\documentclass[12pt,a4paper]{article}
\usepackage{epsfig}
\begin{document}
\title{Production of mirror fermions via $e\gamma$ and $ep$ collisions
in the littlest Higgs model with T-parity}
\author{Chong-Xing Yue, Li Ding, Jin-Yan Liu\\
{\small Department of Physics, Liaoning  Normal University, Dalian
116029, P. R. China}
\thanks{E-mail:cxyue@lnnu.edu.cn}}
\date{\today}

\maketitle
\begin{abstract}

\vspace{1cm}

One of the important features of the littlest Higgs model with
T-parity, called the $LHT$ model, is that it introduces the mirror
fermions, which are the T-parity partners of the standard model
fermions. In this paper, we discuss production of the mirror quark
associated with mirror neutrino via $e\gamma$ and $ep$ collisions.
We find that, in wide range of the parameter space, the mirror quark
can be copiously produced at the International $e^{+}e^{-}$  Linear
Collider $(ILC)$ and $ep$ collider $(THERA)$ experiments. The
production rates of certain signal events, which are related the
main two-body decay modes of the mirror quark, are also calculated.

\end{abstract}
\newpage
\noindent{\bf 1. Introduction}

The $CERN$ Large Hadron Collider ($LHC$) with a center-of-mass
(c.m.) energy of $14TeV$ will begin operation in 2008 for
breakthrough discoveries in the electroweak-scale physics and in the
new physics beyond the standard model ($SM$). However, the processes
and the detection of new physics at the $LHC$ are extremely
complicated. Therefore, the lepton-lepton and lepton-hadron
colliders with clean environments are required to complement the
$LHC$ in drawing a comprehensive and a high-resolution picture of
the $SM$ and the new physics models.

It has been shown that the collective symmetry breaking mechanism
implemented in little Higgs models provides an interesting solution
to the `` little hierarchy problem  " (for recent review see [1]).
The littlest Higgs ($LH$) model [2] is one of the most economical
and interesting models discussed in the literature. However, the low
energy electroweak precision tests enforce the symmetry breaking
scale $f$ of the $LH$ model to be larger than about $4TeV$ [3],
which induces that the fine tuning between the cut-off scale and the
electroweak scale is needed again. To alleviate this difficulty, a
$Z_{2}$ discrete symmetry, named `` T-parity ", is introduced into
the $LH$ model, which forms the so-called $LHT$ model [4]. In the
$LHT$ model, T-Parity is an exact symmetry, the $SM$ gauge bosons
(T-even) do not mix with the T-odd new gauge bosons, and thus the
electroweak observables are not modified at tree level. Beyond the
tree level, small radiative corrections to the electroweak
observables allow the scale parameter $f$ as low as $500GeV$ [5].
Thus, the $LHT$ model is one of the attractive little Higgs models.

In order to implement T-parity in the fermion sector, one introduces
three doublets of `` mirror quarks " and three doublets of `` mirror
leptons ", which have T-odd parity, transform vectorially under
$SU(2)_{L}$. A first study of the collider phenomenology of the
$LHT$ model was given in [6]. The possible signals of the mirror
fermions have been studied in [7, 8, 9, 10, 11]. In Ref.[12] we have
considered pair production of the mirror leptons in an International
$e^{+}e^{-}$  Linear Collider $(ILC)$  and find that, as long as the
mirror leptons are not too heavy, they can be copiously produced via
the processes $e^{+}e^{-}$ $\rightarrow
$$\bar{L_{i}}L_{j}$ in future $ILC$ experiments. In this paper, we will
consider production of the mirror quark associated with mirror
neutrino via $e\gamma$ and $ep$ collisions and see whether the
mirror fermions can be detected via these collision processes in
future $ILC$ and $THERA$ experiments.

In the rest of this paper, we will give our results in detail. In
section 2, we briefly review the couplings of the mirror quarks,
which are related to our calculation. To discuss the possible
signals of the mirror quarks in following sections, their possible
partial decay widths are also given in this section. Production of
the mirror fermions via $e\gamma$ and $ep$ collisions are studied in
sections 3 and 4, respectively. The relevant phenomenology analysis
in future $ILC$ and $THERA$ experiments are also given in these two
sections. In the last section our conclusions and discussions are
given.

\noindent{\bf 2. The couplings and decays of the mirror quarks}

Similar to the $LH$ model, the $LHT$ model [4] is based on an
$SU(5)/SO(5)$ global symmetry breaking pattern. A subgroup
[$SU(2)\times U(1)]_{1} \times [SU(2)\times U(1)]_{2}$ of the
$SU(5)$ global symmetry is gauged, and at the scale $f$ it is broken
into the $SM$ electroweak symmetry $SU(2)_{L} \times U(1)_{Y}$.
T-parity is an automorphism that exchanges the $[SU(2) \times
U(1)]_{1}$ and $[SU(2) \times U(1)]_{2}$ gauge symmetries. The
T-even combinations of the gauge fields are the $SM$ electroweak
gauge bosons $W^{a}_{\mu}$ and $B_{\mu}$. The T-odd combinations are
T-parity partners of the $SM$ electroweak gauge bosons.

To avoid severe constraints and simultaneously implement T-parity,
one needs to double the $SM$ fermion doublet spectrum [4, 6]. The
T-even combination is associated with the $SM$ $SU(2)_{L}$ doublet,
while the T-odd combination is its T-parity partner. To generate
mass terms for the T-odd fermions, called the mirror fermions,
through Yukawa interactions, one requires additional T-odd $SU(2)$
singlet fermions in the $LHT$ model as suggested in [5, 6]. Assuming
universal and flavor diagonal Yukawa coupling $k$, the masses of the
up- and down- type mirror fermions can be written as [11]:
\begin{equation}
M_{U_{H}}\approx\sqrt{2}kf(1-\frac{\nu^{2}}{8f^{2}}),\hspace{0.5cm}
M_{D_{H}}\approx\sqrt{2}kf.
\end{equation}
Being $f\geq500GeV$, it is clear from Eq.(1) that there is
$M_{U_{H}} \approx M_{D_{H}} = M_{Q_{H}}$. In this paper we will
focus our attention on the first and second generation mirror quarks
and assume that the Yukawa coupling constant $k$ is in the range of
$0.5< k < 1.5$.

The couplings of the mirror fermions to other particles, which are
related to our analysis, are summarized as [9]:
\begin{eqnarray}
&&B_{H}\bar{U}_{H}^{i}u^{j}
\hspace{0.1cm}:\hspace{0.1cm}-\frac{ie}{2}[\frac{1}{5C_{W}}+\frac{x_{H}}
{S_{W}}\frac{\nu^{2}}{f^{2}}](V_{Hu})_{ij}\gamma^{\mu}P_{L}, \\
&&B_{H}\bar{U}_{H}^{i}t\hspace{0.3cm} :\hspace{0.1cm}
\frac{ie}{2}[-\frac{1}{5C_{W}}+(\frac{x^{2}_{L}}{10C_{W}}-\frac{x_{H}}
{S_{W}})\frac{\nu^{2}}{f^{2}}](V_{Hu})_{i3}\gamma^{\mu}P_{L},\\
&&B_{H}\bar{D}_{H}^{i}d^{j}\hspace{0.06cm}:\hspace{0.1cm}
-\frac{ie}{2}[\frac{1}{5C_{W}}-\frac{1}{S_{W}}x_{H}\frac{\nu^{2}}{f^{2}}](V_{Hd})_{ij}\gamma^{\mu}P_{L};\\
 &&Z_{H}\bar{U}^{i}_{H}u^{j}\hspace{0.16cm}:\hspace{0.1cm}
\frac{ie}{2}[\frac{1}{S_{W}}-\frac{1}{5C_{W}}x_{H}\frac{\nu^{2}}{f^{2}}](V_{Hu})_{ij}\gamma^{\mu}P_{L},\\
 &&Z_{H}\bar{U}^{i}_{H}t \hspace{0.34cm}:\hspace{0.1cm}
\frac{ie}{2}[\frac{1}{S_{W}}-(\frac{x_{L}^{2}}{2S_{W}}+\frac{x_{H}}{5C_{W}})\frac{\nu^{2}}{f^{2}}]
(V_{Hu})_{i3}\gamma^{\mu}P_{L},\\
 &&Z_{H}\bar{D}^{i}_{H}d^{j}\hspace{0.11cm}:\hspace{0.1cm}
-\frac{ie}{2}[\frac{1}{S_{W}}+\frac{1}{5C_{W}}x_{H}\frac{\nu^{2}}{f^{2}}](V_{Hd})_{ij}\gamma^{\mu}P_{L};\\
&& W_{H}^{\mu}\bar{U}^{i}_{H}d^{j}\hspace{0.08cm}:\hspace{0.1cm}
\frac{ie}{\sqrt{2}S_{W}}(V_{Hd})_{ij}\gamma^{\mu}P_{L},\hspace{2.4cm}
 \hspace{0.1cm}W_{H}^{\mu}\bar{D}^{i}_{H}u^{j}\hspace{0.1cm}:\hspace{0.1cm}
\frac{ie}{\sqrt{2}S_{W}}(V_{Hu})_{ij}\gamma^{\mu}P_{L},\\
 &&W_{H}^{\mu}\bar{D}^{i}_{H}t\hspace{0.19cm}:\hspace{0.1cm}
\frac{ie}{\sqrt{2}S_{W}}(1-\frac{x_{L}^{2}}{2}\frac{\nu^{2}}{f^{2}})(V_{Hu})_{i3}\gamma^{\mu}P_{L},\hspace{0.4cm}
 W_{H}^{\mu}\bar{\nu}^{i}_{H}e\hspace*{0.39cm}:\hspace{0.1cm}
\frac{ie}{\sqrt{2}S_{W}}(V_{H\ell})_{i1}\gamma^{\mu}P_{L}.
 \end{eqnarray}
Here $\nu\approx246GeV$ is the electroweak scale and
$x_{H}=5S_{W}C_{W}/4(5C_{W}^{2}-S_{W}^{2})$. $S_{W}=\sin\theta_{W}$,
 $C_{W}=cos\theta_{W}$, and $\theta_{W}$ is the Weinberg angle. $P_{L}=(1-\gamma_{5})/2$
 is the left-handed projection operator. The four $CKM$-$like$ unitary matrices $V_{Hu}$, $V_{Hd}$,
$V_{H\ell}$, and $V_{H\nu}$ satisfy [9, 13]:
\begin{eqnarray}
  V_{Hu}^{+}V_{Hd}=V_{CKM}, \hspace{0.5cm} V_{H\nu}^{+}V_{H\ell}=V_{PMNS},
\end{eqnarray}
where the $CKM$ matrix $V_{CKM}$ is defined through flavor mixing in
the down-type quark sector, while the $PMNS$ matrix $V_{PMNS}$ is
defined through neutrino mixing. To avoid any additional parameters
introduced and to simply our calculation, we take
$V_{H\ell}=V_{PMNS}$ $(V_{H\nu}=I)$ [14], which means that the
mirror leptons have no impact on the flavor violating observables in
the neutrino sector. For the $CKM$-$like$ unitary matrices $V_{Hu}$
and $V_{Hd}$, we take $V_{Hd}=I$ and $V_{Hu}=V_{CKM}^{+}$ [9] in our
numerical estimation.

At the order of $\nu^{2}/f^{2}$, the masses of the T-odd set of the
$SU(2)\times U(1)$ gauge bosons are given by [4]:
\begin{eqnarray}
M_{B_{H}}=\frac{ef}{\sqrt{5}C_{W}}[1-\frac{5\nu^{2}}{8f^{2}}],\hspace{0.5cm}
M_{Z_{H}}\approx
M_{W_{H}}=\frac{ef}{S_{W}}[1-\frac{\nu^{2}}{8f^{2}}].
\end{eqnarray}
Comparing Eq.(1) with Eq.(11) we can see that, for $0.5\leq k
\leq1.5$, the mirror quarks $U_{H}$ and $D_{H}$ are always heavier
than the T-odd gauge bosons. The gauge boson $B_{H}$ is the lightest
T-odd particle, which can be seen as an attractive dark matter
candidate [6]. Thus, the possible two-body decay modes of the mirror
quarks $U_{H}$ and $D_{H}$ are:
\begin{eqnarray}
U_{H}&:& uB_{H}, \hspace{0.2cm}cB_{H}, \hspace{0.2cm}tB_{H},
\hspace{0.2cm}uZ_{H},
 \hspace{0.2cm}cZ_{H}, \hspace{0.2cm}tZ_{H}, \hspace{0.2cm}dW_{H},\\
D_{H}&:& dB_{H},\hspace{0.2cm} dZ_{H},\hspace{0.2cm}
uW_{H},\hspace{0.2cm} cW_{H}, \hspace{0.2cm}tW_{H}.
\end{eqnarray}
\vspace{-1.005cm}
\begin{figure}[htb]
\begin{center}
\vspace{0.2cm}
 \epsfig{file=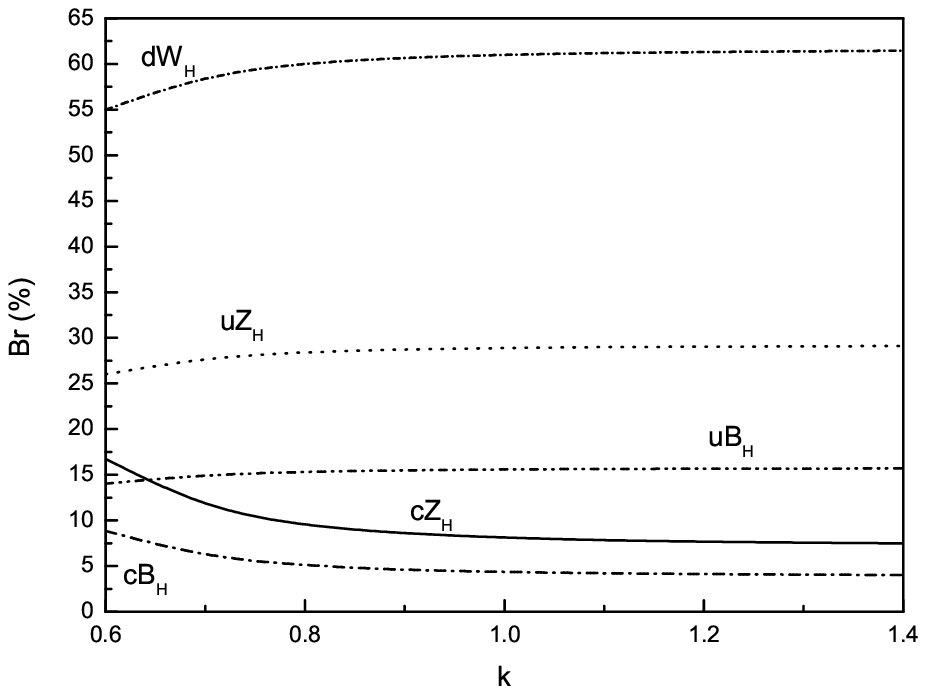,width=225pt,height=210pt}
\put(-110,3){ (a)}\put(115,3){ (b)}
 \hspace{0cm}\vspace{-0.25cm}
\epsfig{file=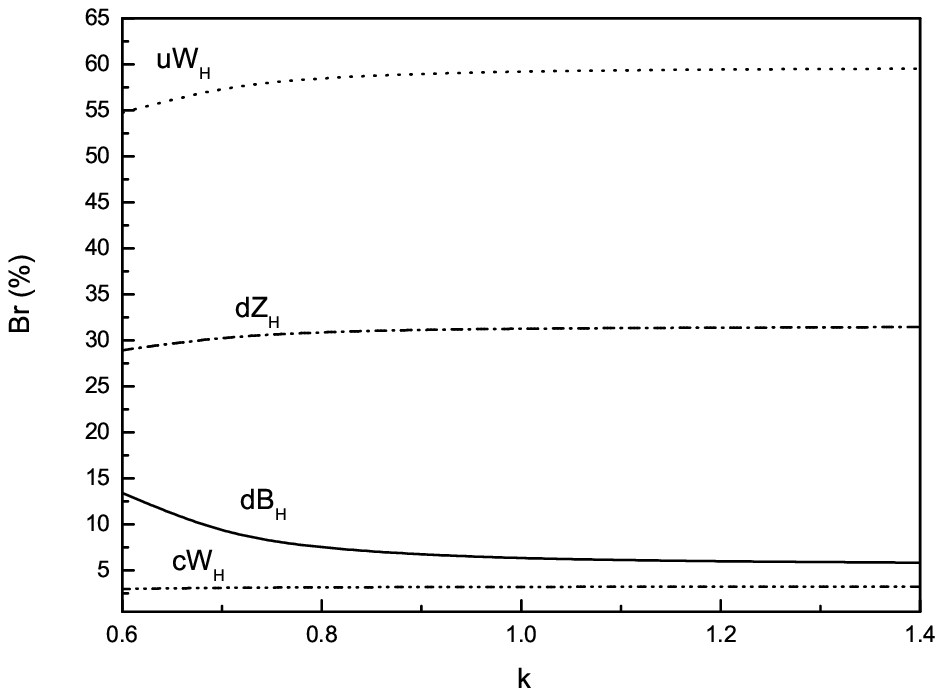,width=225pt,height=210pt} \hspace{-0.5cm}
 \hspace{10cm}\vspace{-1cm}
 \caption{The branching ratios of the mirror quarks $U_{H}$ (a) and
 $D_{H}$ (b) as functions \hspace*{1.9cm}of the coupling parameter $k$ for the scale parameter
 $ f=1TeV $.}
 \label{ee}
\end{center}
\end{figure}

From Eq.(2) -- Eq.(9) we can see that the couplings of the mirror
quark $Q_{H}$ to ordinary fermion and T-odd gauge boson are all the
left-handed couplings. Then the partial decay width can be written
in an unified manner as :
\begin{eqnarray}
\Gamma(Q_{H}\rightarrow qV_{H})=\frac{M_{Q_{H}}^{3}g_{L}^{2}}{32\pi
M_{V_{H}}}\{x^{2}(1-2x^{2}+y^{2})+(1-y^{2})^{2}\}\lambda^{1/2}(1,x^{2},
y^{2})
\end{eqnarray}
with $x=M_{V_{H}}/M_{Q_{H}}$, $y=M_{q}/M_{Q_{H}}$, and
$\lambda(x,y,z)=x^{2}+y^{2}+z^{2}-2xy-2xz-2yz$, in which $M_{V_{H}}$
is the mass of the T-odd gauge boson. Obviously, the branchings
ratios of the mirror quark $Q_{H}$ depend on the free parameters
$f$, $k$, and $x_{L}$. Since the mixing parameter $x_{L}$
contributes the decay widths and the production cross sections
relevant for the top quark $t$ only at the order of $\nu^{4}/f^{4}$,
we will take $x_{L}=0.5$ in our following numerical calculation.

In Fig.1, we illustrate the sizes of the branching ratios for the
mirror quarks $U_{H}$
 and $D_{H}$ as functions of the coupling parameter $k$ for the scale
 parameter $f=1TeV$. Since the values of the branching ratios
 $Br(U_{H} \rightarrow tB_{H})$, $Br(U_{H} \rightarrow tZ_{H})$, and
 $Br(D_{H} \rightarrow tW_{H})$ are all smaller than $2 \times
 10^{-3}\%$, we have not given the relevant curves in Fig.1.
 Furthermore, we have multiplied the factor $10$ to the branching ratio
 $Br(U_{H}\rightarrow c B_{H})$ in Fig.1a. One can see that the
 main decay channels are $uZ_{H}$ and $dW_{H}$ ($dZ_{H}$ and
 $uW_{H}$) for $U_{H}$ ($D_{H}$). For $f=1TeV$ and
 $0.6\leq k\leq1.2$, the values of $Br(U_{H}\rightarrow
 dW_{H})$, $Br(U_{H}\rightarrow uZ_{H})$, $Br(D_{H}\rightarrow
 uW_{H})$, and $Br(D_{H}\rightarrow
 dZ_{H})$ are in the ranges of $55\%\sim61.3\%$, $26\%\sim29\%$,
 $54.7\%\sim59.4\%$, and $28.9\%\sim31.4\%$,
 respectively.

 \noindent{\bf 3. Production of the mirror quark associated with a mirror neutrino
 via $\mathbf{ep}$ \hspace*{0.7cm}collision}

From the above discussion we can see that the mirror quark can be
produced associated with a mirror neutrino via the process $e
q\rightarrow \nu_{H} Q_{H}$ mediated by the T-odd charged gauge
boson $W_{H}$, as depicted in Fig.2. The invariant amplitude for the
process $e(P_{1})q(P_{2})\rightarrow
\nu_{H}^{i}(P_{3})Q_{H}^{j}(P_{4})$ can be written as:
\begin{eqnarray}
M_{q}^{ij}=\frac{e^{2}}{2S_{W}^{2}}\frac{V_{ij}(V_{H\ell})_{1i}}{K^{2}-M_{W_{H}}^{2}}
[\bar{u}(P_{4})\gamma_{\nu}P_{L}u(P_{2})][\bar{u}(P_{3})\gamma^{\nu}P_{L}u(P_{1})],
\end{eqnarray}
where $K^{2}=(P_{4}-P_{2})^{2}$ and $q$ represents the $SM$ light
quark $u$, $c$, $d$, or $s$. $i$ and $j$ are the family indexes for
the mirror fermions. In this paper, we are only interested in
production of the first and second generation mirror fermions, so we
will take $i$=1, 2 and $j$=1, 2. \vspace*{-0.1cm}

\begin{figure}[htb]
\begin{center}
\hspace*{-4.5cm}\epsfig{file=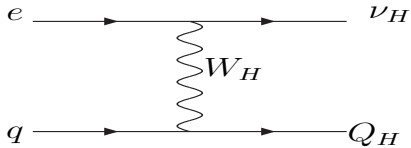,width=700pt,height=700pt}\vspace*{-21.5cm}
 \caption{Feynman diagram for the process $eq\rightarrow\nu_{H} Q_{H}$.}
 \label{ee}
\end{center}
\end{figure}

After calculating the cross section $\hat{\sigma}_{q}^{ij}(\hat{s})$
for the process $e(P_{1})q(P_{2})\rightarrow
\nu_{H}^{i}(P_{3})Q_{H}^{j}(P_{4})$, the effective production cross
section $\sigma(s)$ of the mirror fermions at the $THERA$ experiment
with the center-of-mass (c.m.) energy $\sqrt{S}=3.7TeV$ [15] can be
obtained by folding $\hat{\sigma}_{q}^{ij}(\hat{s})$ with the parton
distribution function ($PDF$) $f_{q/p}(x)$:
\begin{eqnarray}
\sigma(s)=\sum_{i, j}\sum_{q}\int^{1}_{x_{min}}f_{q/p}(x,
\mu_{F})\hat{\sigma_{q}}^{ij}(\hat{s})dx
\end{eqnarray}
with $x_{min}=(M_\nu+M_{Q_{H}})^{2}/s$ and $\hat{s}=xs$, in which
$M_\nu $ is the mass of the mirror neutrino $ \nu_{H}$. In our
numerical calculation, we will use $CTEQ6L$ $PDF$ [16] for the quark
distribution function $f_{q/p}(x, \mu_{F})$ and assume that the
factorization scale $\mu_{F}$ is of order $\sqrt{\hat{s}}$.

\begin{figure}[htb]
\begin{center}
\vspace{-1.3cm}
 \epsfig{file=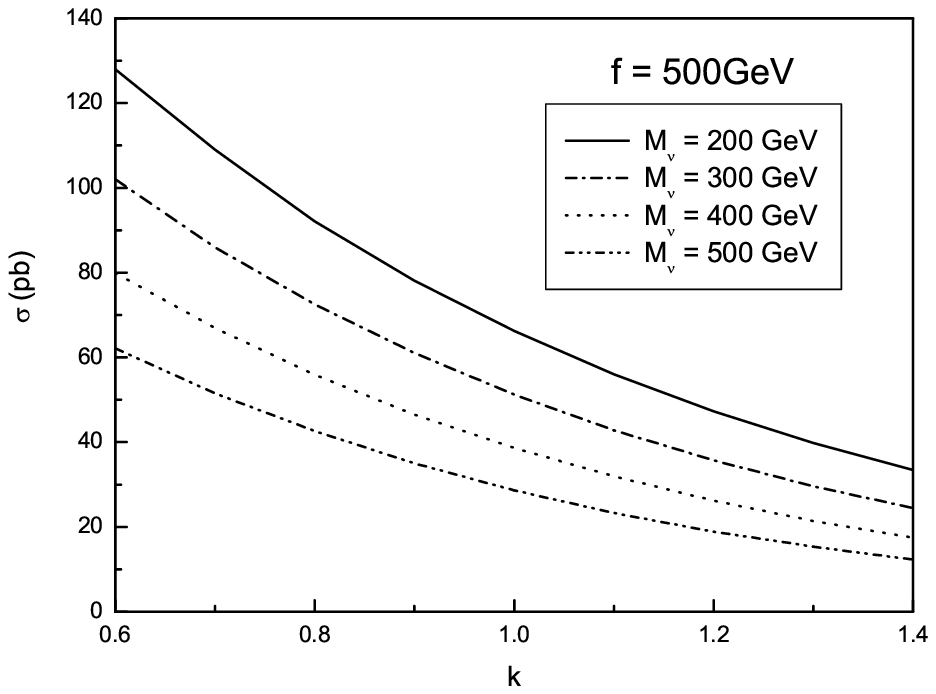,width=225pt,height=210pt}
\put(-110,3){ (a)}\put(115,3){ (b)}
 \hspace{0cm}\vspace{-0.25cm}
\epsfig{file=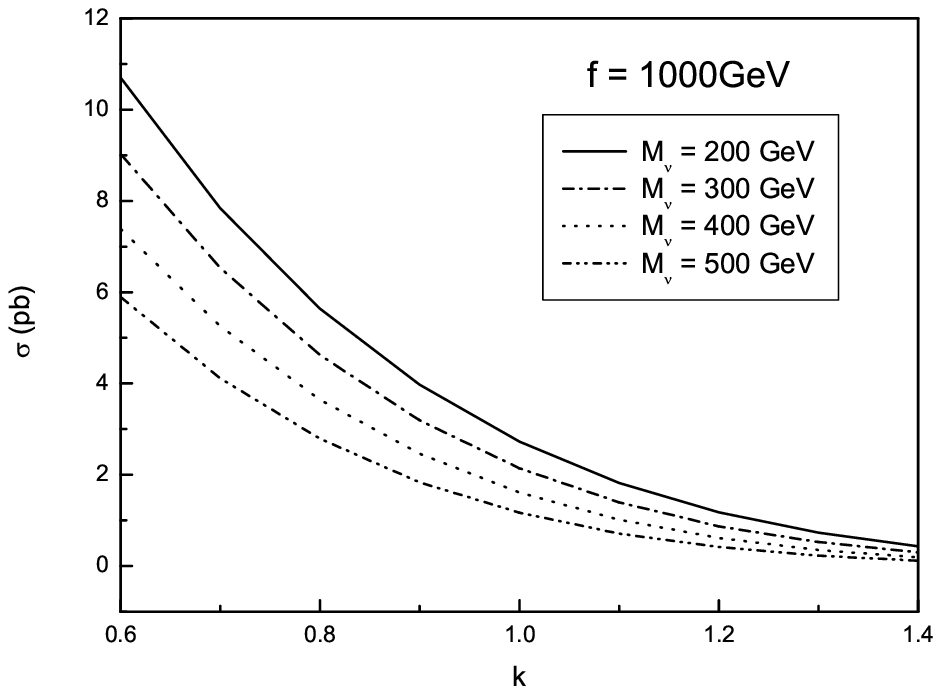,width=225pt,height=210pt} \vspace{0.2cm}
 \epsfig{file=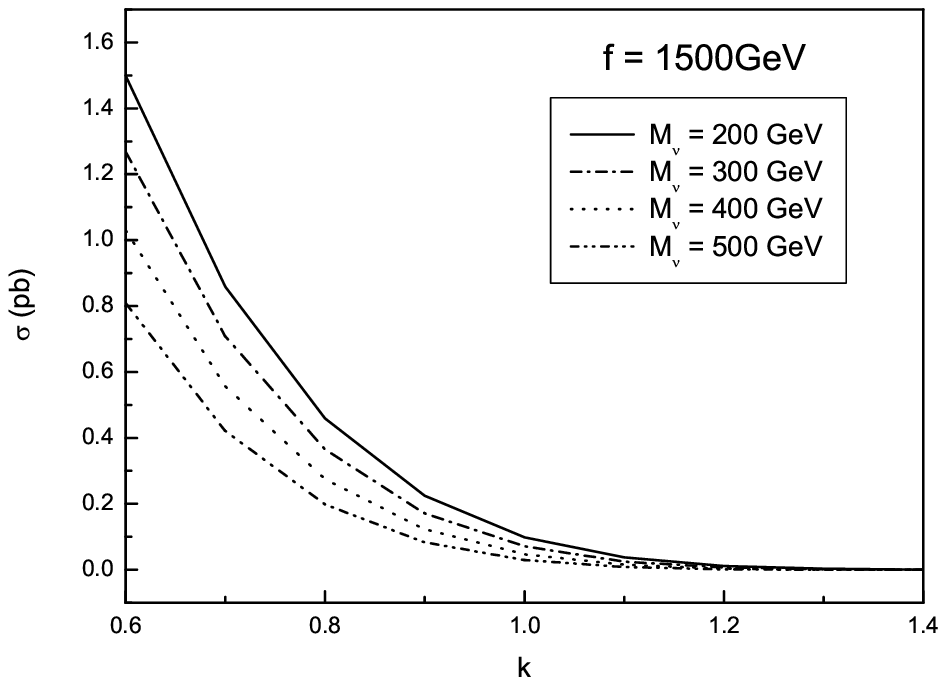,width=225pt,height=210pt}
\put(-110,3){ (c)}\put(115,3){ (d)}
 \hspace{0cm}\vspace{-0.25cm}
\epsfig{file=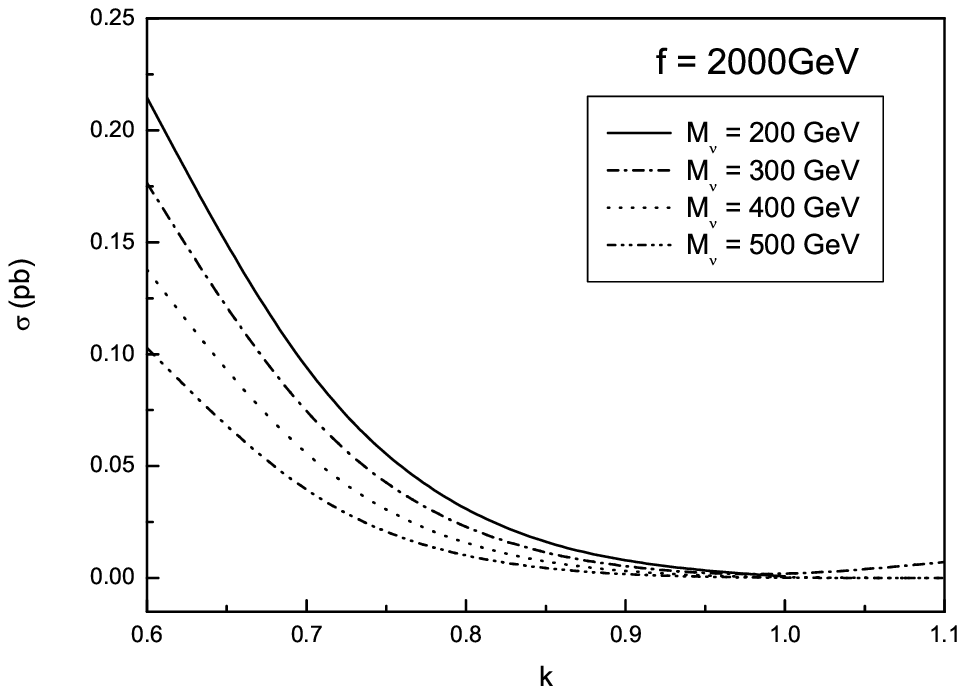,width=225pt,height=210pt} \hspace{-0.5cm}
 \hspace{10cm}\vspace{-1cm}
 \caption{The effective cross sections $\sigma$ for the subprocess
 $eq\rightarrow\nu_{H} Q_{H}$ as a function
 \hspace*{1.8cm} of
the parameter $k$ for different values of the free parameters
$M_{\nu}$ and $f$.}
 \label{ee}
\end{center}
\end{figure}

In the case of $V_{H\ell}=V_{PMNS}$, $V_{Hd}=I$, and
$V_{Hu}=V_{CKM}^{+}$, the production cross section $\sigma(s)$
depends on the parameters $f$, $k$, and $M_{\nu}$. Our numerical
results are given in Fig.3, in which we plot the cross section
$\sigma(s)$ as a function of the coupling parameter $k$ for
different values of the free parameters $f$ and $M_{\nu}$. In this
figure we have taken the values of the $CKM$ matrix elements
$(V_{CKM})_{ij}$ given in [17], in which $V_{CKM}$ is constructed
based on $PDG$ parameterization [18]. For the matrix $V_{PMNS}$, we
have taken the standard parameterization form parameters given by
the neutrino experiments [19]. The production cross section
$\sigma(s)$ given by Fig.3 is the total cross section for the first
two generation up-type and down-type quarks. One can see from Fig.3
that the value of $\sigma(s)$ decreases as the parameters $f$, $k$,
and $M_{\nu}$ increasing. For $M_{\nu}=300GeV$, $k=1.0$, and $500GeV
\leq f \leq 2000GeV$, its value is in the range of
$51.2pb\sim3.4\times10^{-4}pb$. For $f\geq1500GeV$ and $k \sim 1$,
the mirror fermion is too heavy to be abundantly produced at the
$THERA$ experiment with $\sqrt{S}=3.7TeV$, and its production cross
sections is smaller than $98fb$, which is very difficult to be
detected in future $THERA$ experiments.

For $k>0.5$, the mirror quark is heavy enough to decay into T-odd
gauge boson plus an ordinary fermion. The different chain decays of
the mirror quark can given different experimental signatures. To
asses the discovery potential of the $THERA$ for the mirror quarks,
the production rates and the relevant $SM$ backgrounds for different
decay channels need to be discussed.

The decay processes $\nu_{H} U_{H}\rightarrow \nu_{H} u B_{H}$,
$\nu_{H} c B_{H}$ and $\nu_{H} D_{H}\rightarrow \nu_{H} d B_{H}$ can
lead to the $jet+E$ \hspace{-0.55cm} / signature. To illustrate the
size of the signal production rate, we show in Fig.4 it as a
function of the parameter $k$ for different values of the mass
parameter $M_{\nu}$ and scale parameter $f$. One can see from Fig.4
that, for $M_{\nu}=200GeV$, $k=1$ and $f \geq 500GeV$, the
production rate can reach $5.2pb$. The $SM$ backgrounds of this kind
of signature mainly come from the process $ep \rightarrow\nu q'+X$.
Measurement and $QCD$ analysis of the production cross section for
the $SM$ process $ep\rightarrow\nu q'+X$ at the $HERA$ collider have
been extensively studied [20]. In order to see whether the $jet+ E
\hspace{-0.25cm} /$   signal can be detected in future $THERA$
experiment with the $c.$ $m.$ energy $\sqrt{S}=3.7TeV$ and a yearly
integrated luminosity $\pounds =4fb^{-1}$, we further calculate its
statistical significance $SS$, which is defined as:
\begin{eqnarray}
SS=\frac{\sigma(signal)}{\sqrt{\sigma(SM)}}\sqrt{\pounds}.\nonumber
\end{eqnarray}
\begin{figure}[htb]
\begin{center}
\vspace{0.3cm}
 \epsfig{file=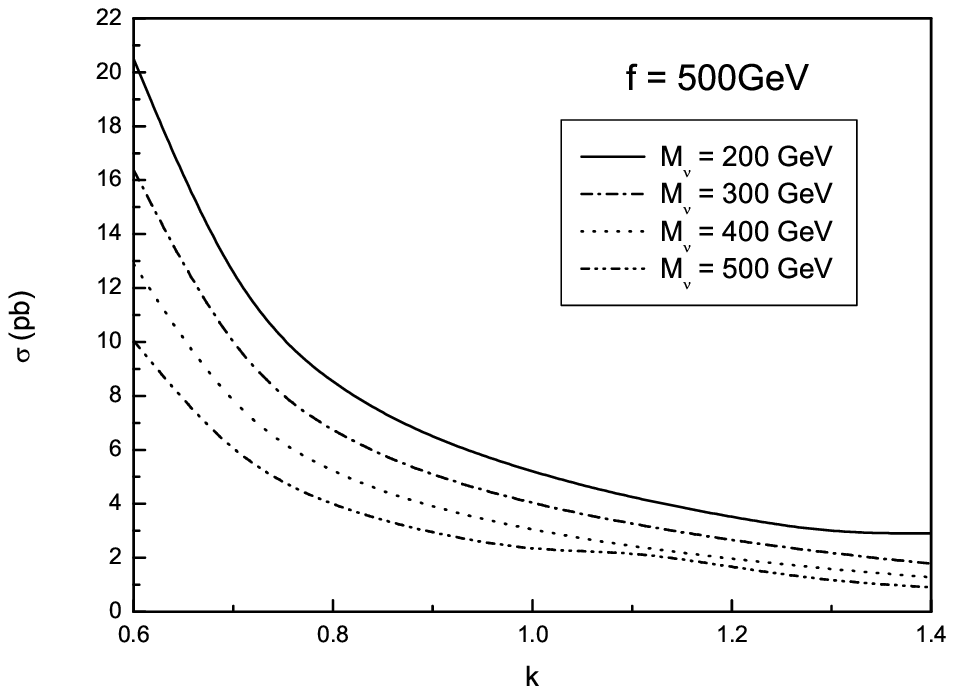,width=225pt,height=210pt}
\put(-110,3){ (a)}\put(115,3){ (b)}
 \hspace{0cm}\vspace{-0.25cm}
\epsfig{file=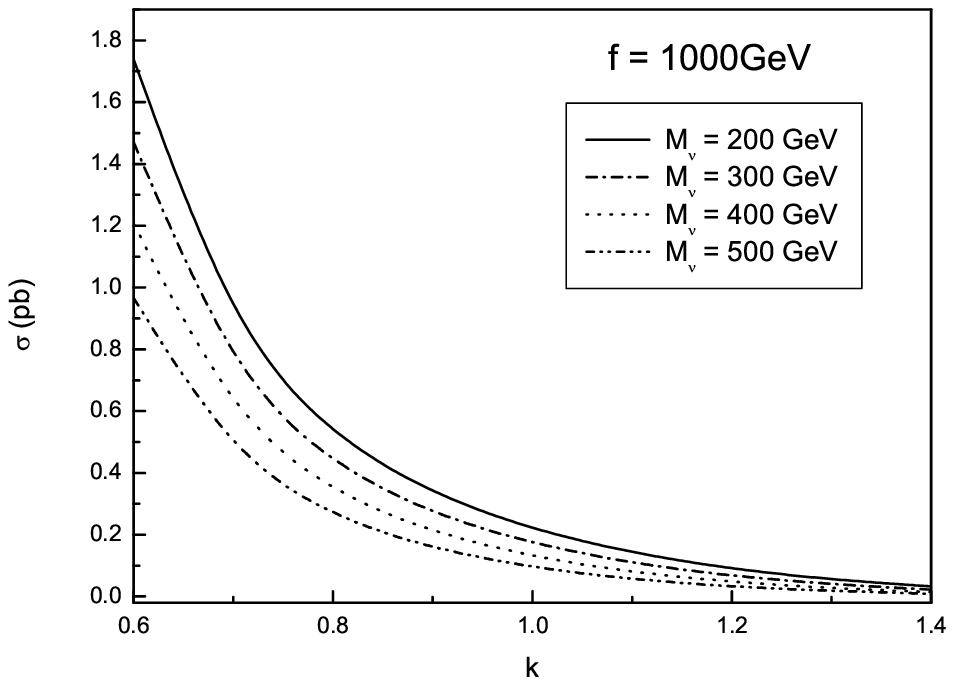,width=225pt,height=210pt} \vspace{0.2cm}
 \epsfig{file=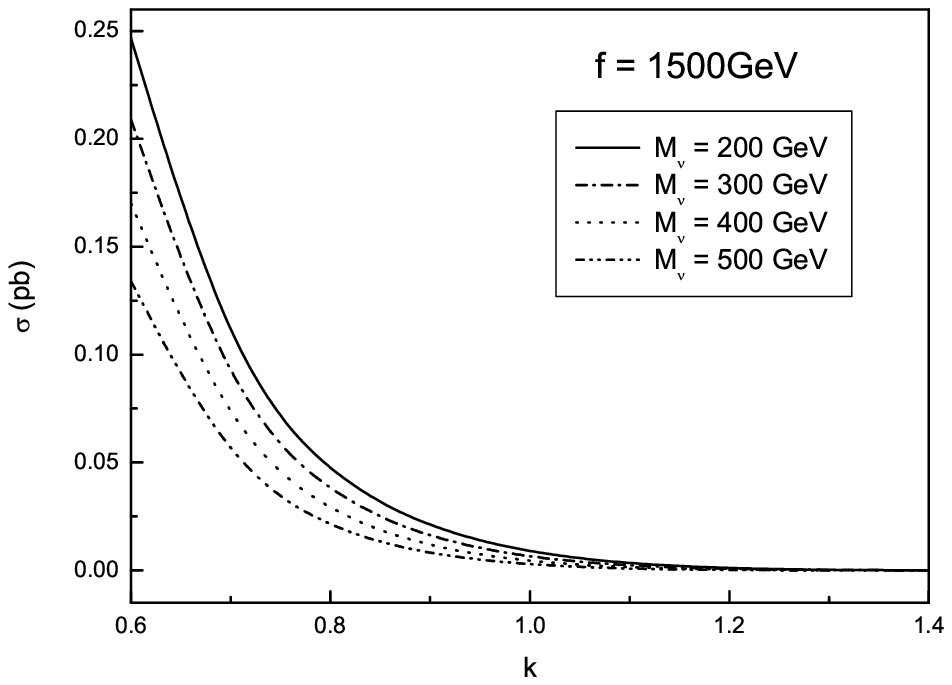,width=225pt,height=210pt}
\put(-110,3){ (c)}\put(115,3){ (d)}
 \hspace{0cm}\vspace{-0.25cm}
\epsfig{file=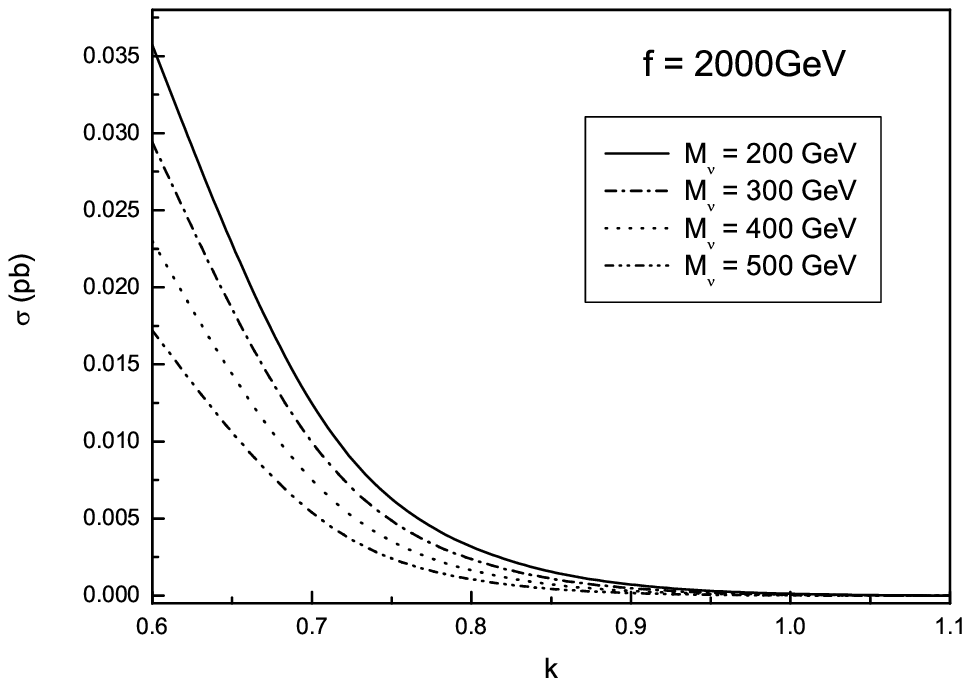,width=225pt,height=210pt} \hspace{-0.5cm}
 \hspace{10cm}\vspace{-1cm}
 \caption{The production rate for the $jet+ E \hspace{-0.25cm} /$
 signature as a function  of the parameter \hspace*{1.8cm} $k$ for different values of the free parameters
$M_{\nu}$ and $f$.}
 \label{ee}
\end{center}
\end{figure}
\vspace{0cm}

To show the sensitivity of the statistical significance $SS$ to the
scale parameter $f$, we plot in Fig.5 $SS$ as a function of $f $ for
a fixed value $M_{\nu}=300GeV$ and three values of the coupling
parameter $k$. By varying $f$ from 500GeV to 1300GeV, the value of
$SS$ is in the rang of $196 \sim 2.9$ for $k=0.8$. Thus, we expect
that, in wide range of the parameter space, the possible signals of
the mirror quark can be detected via the process $ep \rightarrow
\nu_{H}Q_{H}+X$ in future $THERA$ experiments.

\begin{figure}[htb]
\begin{center}
 \epsfig{file=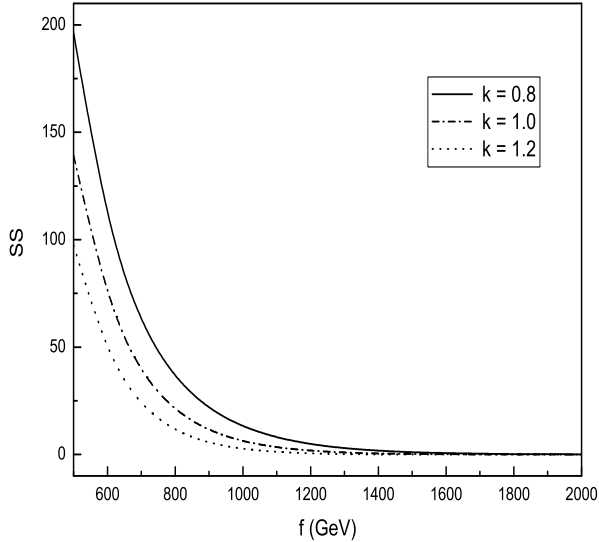,width=250pt,height=240pt}
 \caption{The statistical significance $SS$ as a function of the parameter $f$ for
 $M_{\nu}=\hspace*{2.0cm}300GeV$ and three values of the coupling parameter $k$.}
 \label{ee}
\end{center}
\end{figure}

For the decay channel $U_{H}\rightarrow tB_{H}$, the process $e
p\rightarrow\nu_{H} U_{H}+X$ gives rise to the signal event with a
single top and large missing energy ($t+E$ \hspace{-0.5cm} / ),
which can give a characteristic signal. However, its production rate
is smaller than $1fb$ and can not be detected in future $THERA$
experiments [21]. Same as $U_{H} \rightarrow tB_{H}$, the decay
channel $U_{H}\rightarrow t Z_{H}$ can also generate a signal at
unobservable level in future $THERA$ experiments.
\begin{figure}[htb]
\begin{center}
\vspace{-0.5cm}
 \epsfig{file=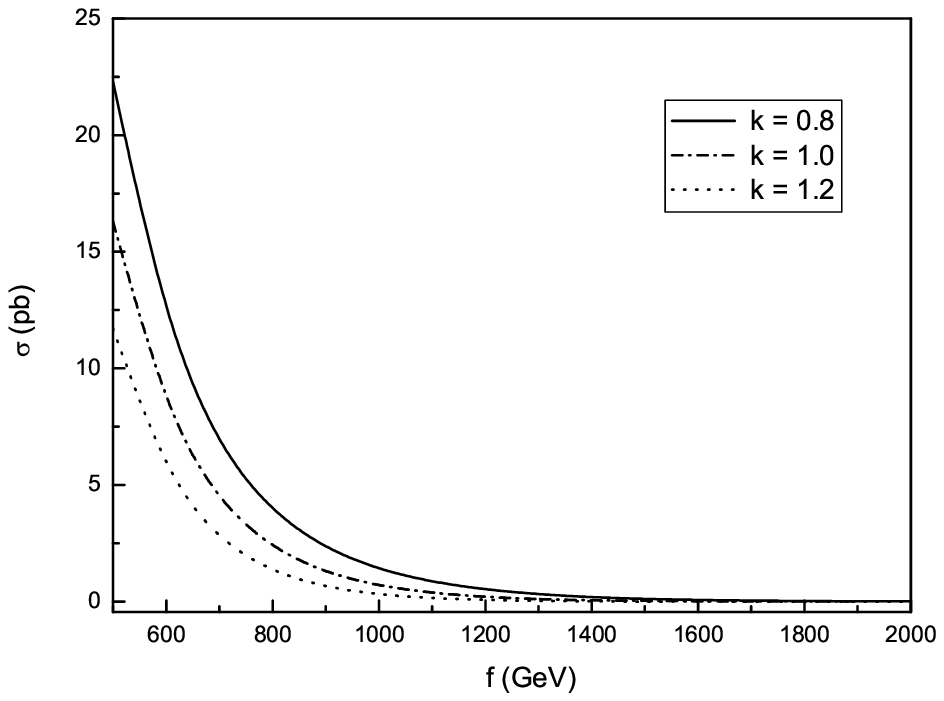,width=225pt,height=210pt}
\put(-110,3){ (a)}\put(115,3){ (b)}
 \hspace{0cm}\vspace{-0.25cm}
\epsfig{file=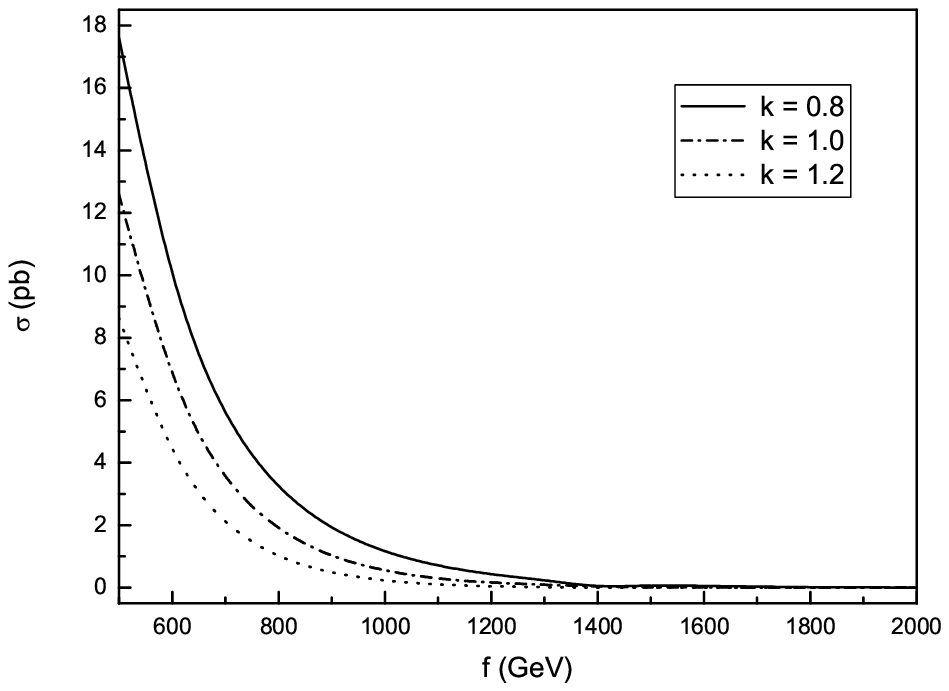,width=225pt,height=210pt} \hspace{-0.5cm}
 \hspace{10cm}\vspace{-1cm}
 \caption{The production rate for the $E \hspace{-0.25cm} /$ + $jets$ signature
 as a function of the parameter \hspace*{1.9cm}$f$ for
 $M_{\nu}=200GeV$(a), and $M_{\nu}=300GeV$(b),
 and three values of $k$.}
 \label{ee}
\end{center}
\end{figure}

For the mirror quark decays into the T-odd gauge boson $Z_{H}$ and
an ordinary fermion, we have the following chain decay
processes:
\begin{eqnarray}
\nu_{H} U_{H}\rightarrow \nu_{H} uZ_{H}\rightarrow \nu_{H} u
B_{H}H,\hspace{0.5cm}\nu_{H} U_{H}\rightarrow \nu_{H}
cZ_{H}\rightarrow \nu_{H} cB_{H}H,
\end{eqnarray}
\begin{eqnarray}
\hspace{-6.0cm}\nu_{H} D_{H}\rightarrow \nu_{H} dZ_{H}\rightarrow
\nu_{H} dB_{H}H.
\end{eqnarray}
For $M_{H}\leq120GeV$, the dominant decay channel of the Higgs boson
$H$ is $H\rightarrow b\overline{b}$. The above chain decay processes
can generate the $E \hspace{-0.25cm} /$ + $jets$ signature. Its
production rate is given in Fig.6, in which we have assumed
$Br(Z_{H}\rightarrow B_{H}H)\approx 100 \% $ [7]. For
$M_{\nu}=300GeV$, $k=1$, and $500GeV \leq f \leq1500GeV$, there will
be $5\times10^{4}\sim68$ signal events to be generated per year in
future $THERA$ with $\sqrt{S}=3.7TeV$ and $\pounds=4fb^{-1}$. The
signal events should be distinguished from the $SM$ backgrounds by
reconstructing the final states.

If we assume that the mirror quark decays into the T-odd gauge boson
$W_{H}$ and an ordinary fermion, then the production of the mirror
quark associated a mirror neutrino can give following chain decay
processes:
\begin{eqnarray}
\nu_{H} U_{H}\rightarrow \nu_{H} d W_{H}\rightarrow \nu_{H}
B_{H}dW\rightarrow \nu_{H} B_{H}d\ell\nu_{\ell},
\end{eqnarray}
\begin{eqnarray}
\nu_{H} D_{H}\rightarrow \nu_{H} u W_{H}\rightarrow \nu_{H} u W
B_{H}\rightarrow \nu_{H} B_{H}u \ell\nu_{\ell},
\end{eqnarray}
\begin{eqnarray}
\nu_{H} D_{H}\rightarrow \nu_{H} cW_{H}\rightarrow \nu_{H}
cB_{H}W\rightarrow\nu_{H} B_{H}c\ell\nu_{\ell}.
\end{eqnarray}
In the above processes, we have assumed that the T-odd gauge boson
$W_{H}$ mainly decays into $B_{H}W$ and focus our attention only on
the pure leptonic decay modes for the $SM$ gauge boson $W$. These
processes can give rise the signal event $E \hspace{-0.25cm} /$
+$\ell+jet$, which is shown in Fig.7. The production rate for the
signal $E \hspace{-0.25cm} /$ +$\ell+jet$ is smaller than that for
the signal event $E \hspace{-0.25cm} /$ +$jets$. However, for
$M_{\nu}=300GeV$, $k=0.8$, and $f \geq 500GeV$, its value can reach
$14.3pb$. In wide range of the parameter space, there will be
several tens and up to thousands $E \hspace{-0.25cm} /$ +$\ell+jet$
events to be generated per year in future $THERA$ experiments.
\vspace{0.5cm}
\begin{figure}[htb]
\begin{center}
\vspace{-0.5cm}
 \epsfig{file=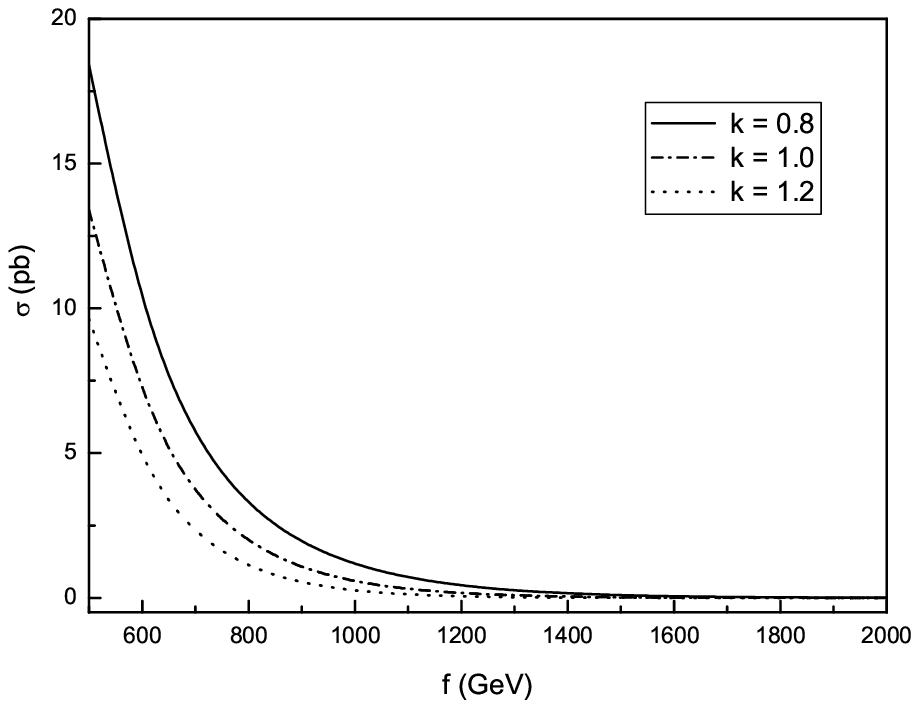,width=225pt,height=210pt}
\put(-110,3){ (a)}\put(115,3){ (b)}
 \hspace{0cm}\vspace{-0.25cm}
\epsfig{file=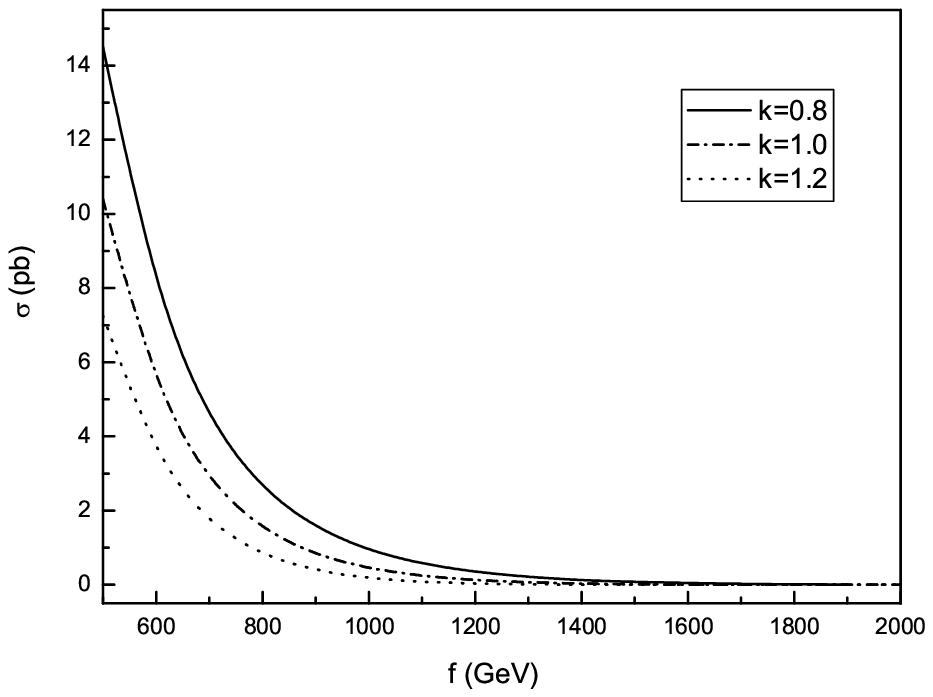,width=225pt,height=210pt} \hspace{-0.5cm}
 \hspace{10cm}\vspace{-1cm}
\caption{The production rate for the signal event $E
\hspace{-0.25cm} /$ +$\ell+jet$
 as a function of the parameter \hspace*{1.9cm}$f$ for
 $M_{\nu}=200GeV$(a), and $M_{\nu}=300GeV$(b),
 and three values of $k$.}
 \label{ee}
\end{center}
\end{figure}

\noindent{\bf 4. Production of the mirror quark associated with a
mirror neutrino via $\mathbf{e\gamma}$ \hspace*{0.6cm} collision}

It has been shown that, in suitable kinematic region the process
$e\gamma\rightarrow \nu q'\bar{q}$ can be approximated quite well by
the process $eq\rightarrow\nu q'$ [22], where the quark $q$
described by the quark parton content of the photon approach [23].
So production of the mirror quark associated with a mirror neutrino
can be induced via $e\gamma$ collision mediated by the T-odd charge
gauge boson $W_{H}$. The hard photon beam of $e\gamma$ collision can
be obtained from laser backscattering at the high energy
$e^{+}e^{-}$ collider experiments. The expression for the total
effective cross section of the subprocess
$e(P_{1})q(P_{2})\rightarrow \nu_{H} ^{i}(P_{3})Q_{H}^{j}(P_{4})$ at
the $ILC$ can be given by
\begin{eqnarray}
\sigma(s)=\sum_{i, j}\sum_{q}\int\int
dx_{1}dx_{2}f_{\gamma/e}(x_{1})f_{q/\gamma}(x_{2})\hat{\sigma_{q}}^{ij}(\hat{s}).
\end{eqnarray}
Here $f_{\gamma/e}(x_{1})$ is the photon distribution function [24],
$f_{q/\gamma}(x_{2})$ is the distribution function for the quark
content in the photon. To obtain our numerical results we will use
$Aurenche$, $Fontannaz$, and $Guillet$ ($AFG$) distributions [25]
for $f_{q/\gamma}$. Other distributions are available in [26].
Furthermore, we will give the cross section for $\nu_{H} Q_{H}$
production at the $ILC$ experiment with $\sqrt{S}=3TeV$ and
$\pounds=100fb^{-1}$ [27].

\begin{figure}[htb]
\begin{center}
\vspace{-0.5cm}
 \epsfig{file=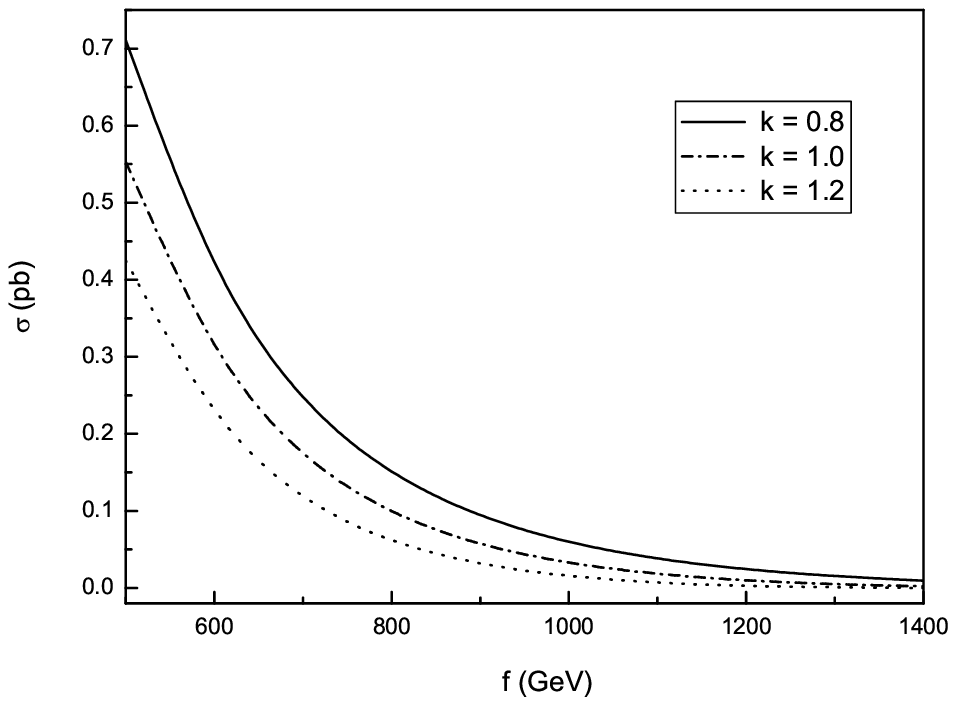,width=225pt,height=210pt}
\put(-110,3){ (a)}\put(115,3){ (b)}
 \hspace{0cm}\vspace{-0.25cm}
\epsfig{file=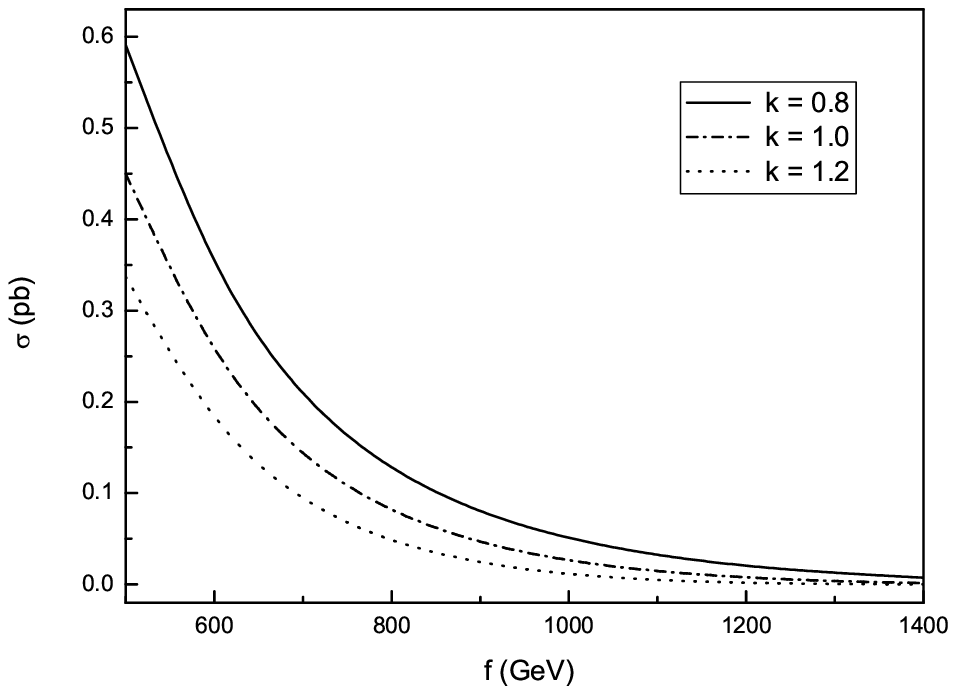,width=225pt,height=210pt} \hspace{-0.5cm}
 \hspace{10cm}\vspace{-1cm}
 \caption{The production cross section for the subprocess
 $eq\rightarrow\nu_{H} Q_{H}$ as a function of \hspace*{1.9cm}the parameter $f$ for
 $M_{\nu}=200GeV$(a), and $M_{\nu}=300GeV$(b),
 and three values \hspace*{1.9cm}of the parameter $k$.}
 \label{ee}
\end{center}
\end{figure}

The production cross section $\sigma$ of the process
$e^{+}e^{-}\rightarrow \nu_{H} Q_{H}+X$ is shown in Fig.8 as a
function of the Yukawa coupling parameter $k$ for different values
of the scale parameter $f$ and the mirror neutrino mass $M_{\nu}$,
in which we have assumed $q=u$, $c$, $d$, and $s$. The cross section
$\sigma$ is the total production cross section of the first two
mirror quarks. From Fig.8 one can see that $\sigma$ is generally
smaller than that for the $THERA$ experiment. The reason is that,
compared with the $PDF$ $f_{q/p}$, the $PDF$'s factor $f_{\gamma/e}$
$f_{q/\gamma}$ suppresses the cross section for the production of
the mirror quark associated with a mirror neutrino. For
$M_{\nu}=200GeV$, $k=1.0$, and $500GeV \leq f \leq 1400GeV$, the
value of $\sigma$ is in the range of $553fb \sim 2.2fb$.
\begin{figure}[htb]
\begin{center}
\vspace{-0.5cm}
 \epsfig{file=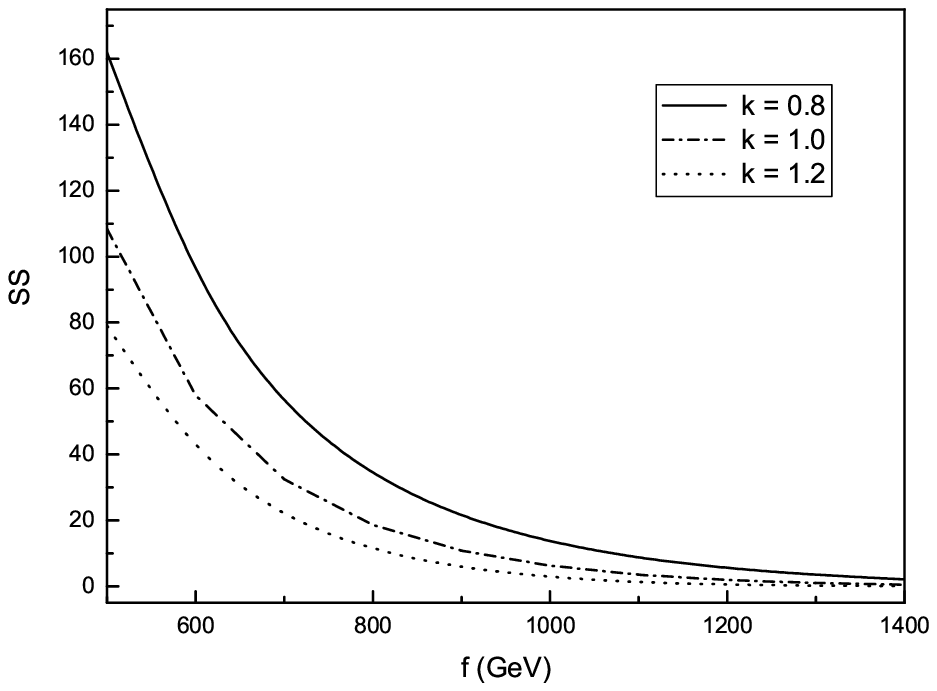,width=225pt,height=210pt}
\put(-110,3){ (a)}\put(115,3){ (b)}
 \hspace{0cm}\vspace{-0.25cm}
\epsfig{file=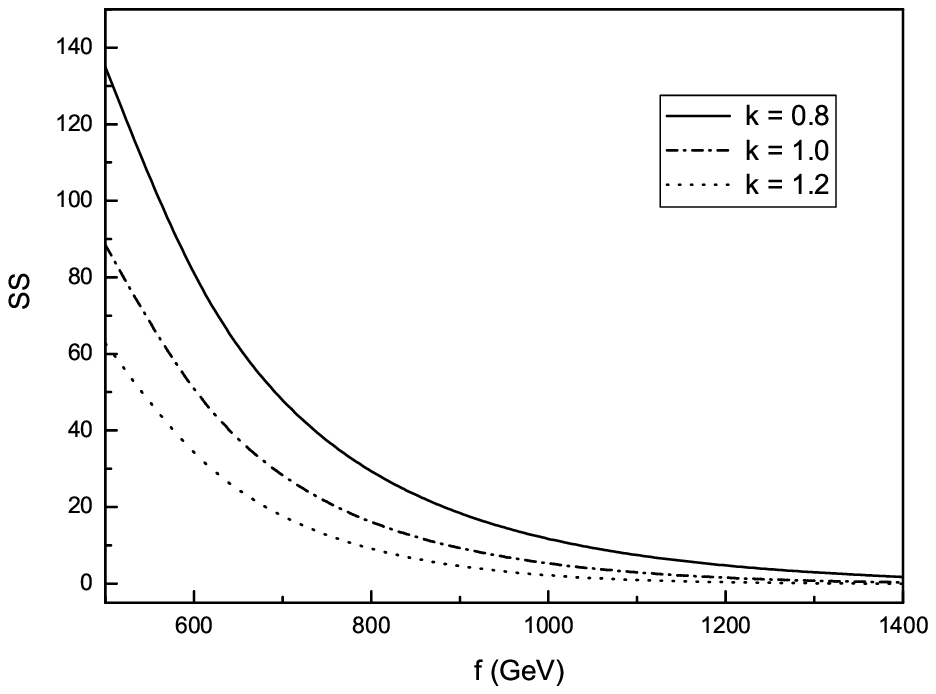,width=225pt,height=210pt} \hspace{-0.5cm}
 \hspace{10cm}\vspace{-1cm}
 \caption{The statistical significance
$SS$ as a function of the parameter $f$ for
 $M_{\nu}= \hspace*{1.9cm}200GeV$(a), and $M_{\nu}=300GeV$(b),
 and three values of the parameter $k$ .}
 \label{ee}
\end{center}
\end{figure}

If we assume that the mirror quarks have the decay channels
$U_{H}\rightarrow uB_{H}$, $U_{H}\rightarrow cB_{H}$, and
$D_{H}\rightarrow dB_{H}$, then the signal state of the process
$e^{+}e^{-}\rightarrow \nu_{H}Q_{H}+X$ is $jet$+$E \hspace{-0.25cm}
/$. The $SM$ backgrounds of this signature mainly come from the $SM$
process $e^{+}e^{-}\rightarrow \nu q+X$. Its statistical
significance $SS$ is plotted as a function of the scale parameter
$f$ for $M_{\nu}=200GeV$ (Fig.9a), $300GeV$ (Fig.9b), and three
values of the coupling parameter $k$. For $k \leq 1.2$, $M_{\nu}
\leq 300GeV$, and $f \leq 1TeV$, the value of the statistical
significance $SS$ is larger than 2. Thus, for reasonable ranges of
the free parameters, the possible signatures of the $LHT$ model
might be detected via the process $e^{+}e^{-} \rightarrow
\nu_{H}Q_{H}+X$ with $Q_{H} \rightarrow qB_{H}$ in future $ILC$
experiments.

For the chain decay processes $U_{H} \rightarrow uZ_{H} \rightarrow
uB_{H}H \rightarrow B_{H}ub\bar{b}$, $U_{H} \rightarrow cZ_{H}
\rightarrow B_{H}cb\bar{b}$, and $D_{H} \rightarrow dZ_{H}
\rightarrow B_{H}db\bar{b}$, the process $e^{+}e^{-} \rightarrow
\nu_{H}Q_{H}+X$ can produce the $E \hspace{-0.25cm} /$+$jets$
signature. In Fig.10 we plot its production rate as a function of
the scale parameter $f$ for different values of the free parameters
$M_{\nu}$ and $k$. One can see from Fig.10 that, for
$M_{\nu}=200GeV$, $k=1.0$, and $500GeV \leq f \leq 1400GeV$, there
will be $1.36\times10^{4} \sim 55$ $E \hspace{-0.25cm} /$+$jets$
events to be generated per year in future $ILC$ experiment with
$\sqrt{S}=3TeV$ and $\pounds=100fb^{-1}$.
\begin{figure}[htb]
\begin{center}
\vspace{-1cm}
 \epsfig{file=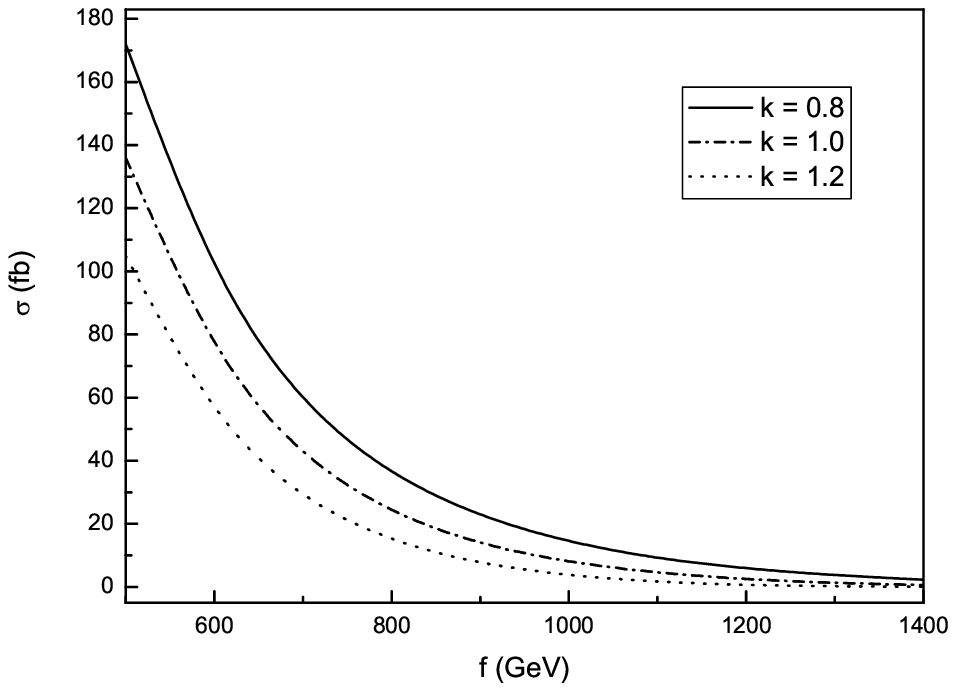,width=225pt,height=210pt}
\put(-110,3){ (a)}\put(115,3){ (b)}
 \hspace{0cm}\vspace{-0.25cm}
\epsfig{file=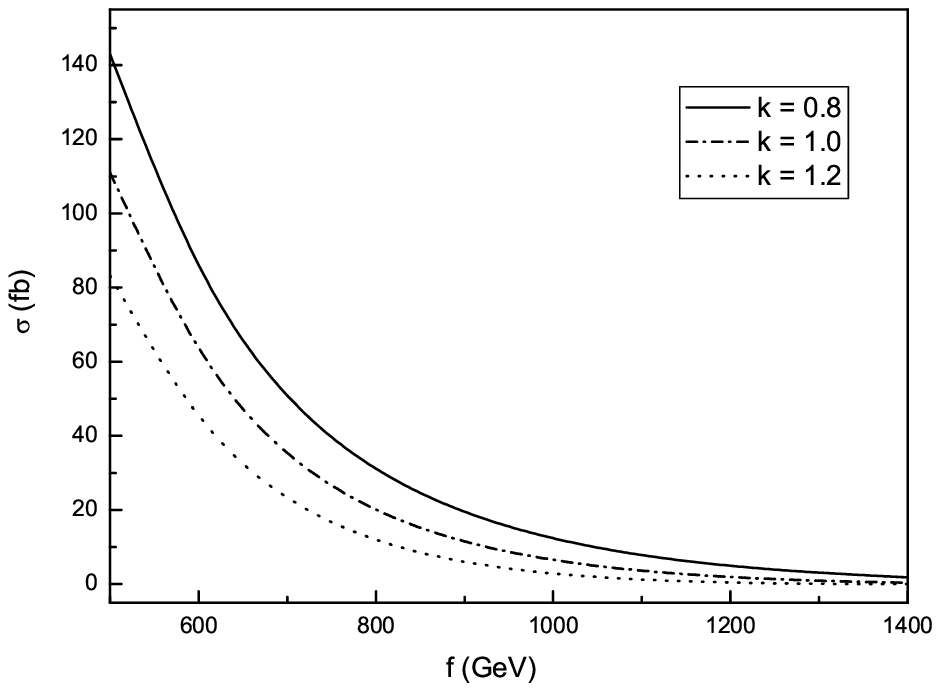,width=225pt,height=210pt} \hspace{-0.5cm}
 \hspace{10cm}\vspace{-1cm}
 \caption{The production rate for the $E \hspace{-0.25cm} /$ + $jets$ signature
 as a function of the parameter \hspace*{1.9cm}$f$ for
 $M_{\nu}=200GeV$(a), and $M_{\nu}=300GeV$(b),
 and three values of $k$.}
 \label{ee}
\end{center}
\end{figure}

\begin{figure}[htb]
\begin{center}
\vspace{-1.6cm}
 \epsfig{file=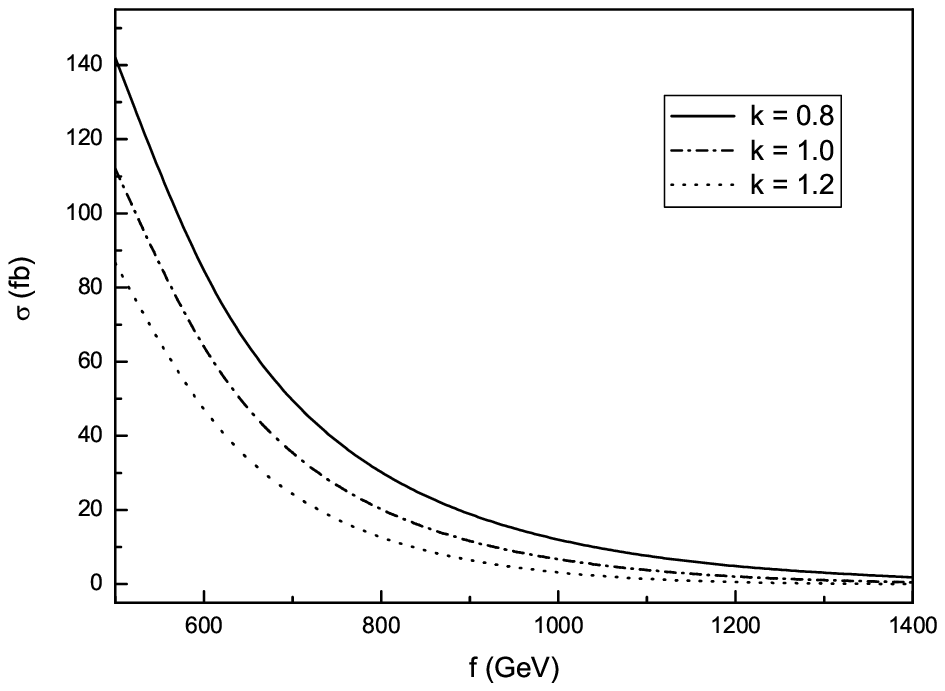,width=225pt,height=210pt}
\put(-110,3){ (a)}\put(115,3){ (b)}
 \hspace{0cm}\vspace{-0.25cm}
\epsfig{file=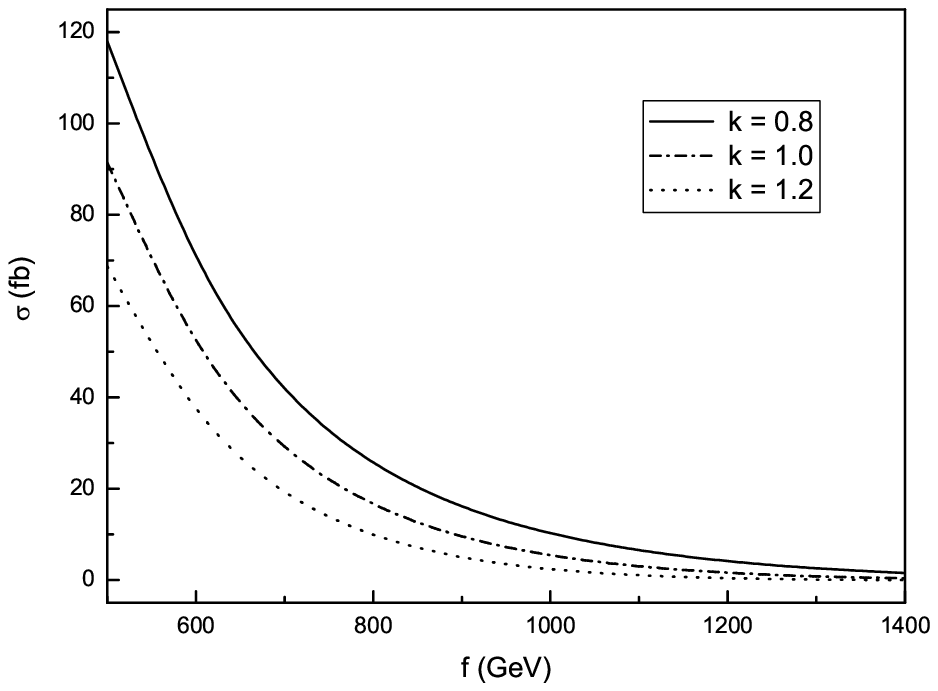,width=225pt,height=210pt} \hspace{-0.5cm}
 \hspace{10cm}\vspace{-1cm}
\caption{The production rate for the signal event $E
\hspace{-0.25cm} /$ +$\ell+jet$
 as the parameter of $f$ for
 \hspace*{1.9cm}$M_{\nu}=200GeV$(a), and $M_{\nu}=300GeV$(b),
 and three values of $k$.}
 \label{ee}
\end{center}
\end{figure}

According the chain decay processes Eq.(19) -- Eq.(21), the process
$e^{+}e^{-} \rightarrow \nu_{H}Q_{H}+X$ can induce the $E
\hspace{-0.25cm} /$ +$\ell+jet$ final state. Our numerical results
show that this signal event can also be abundantly produced in
future $ILC$ experiment with $\sqrt{S}=3TeV$ and
$\pounds=100fb^{-1}$, as illustrated in Fig.11. For
$M_{\nu}=200GeV$, $k=1.0$ and $500GeV \leq f \leq 1400GeV$, there
will be $1.12\times10^{4} \sim 45$ $E \hspace{-0.25cm} /$
+$\ell+jet$ events to be generated per year.

\noindent{\bf 5. Conclusion}

The $LHT$ model is one of the attractive little Higgs models which
not only provides a possible dark matter candidate but also is
consistent with electroweak precision tests. In order to implement
T-parity in the fermion sector of the model, the heavy T-odd $SU(2)$
doublet fermions, which are called the mirror fermions of the $SM$
fermions, have to be introduced. These new heavy fermions might
produce the observability signatures in future high energy collider
experiments.

Recently, we have discussed the possibility of detecting the new
charged gauge boson $W'$ via the process $eq \rightarrow \nu q'$ and
find that one can use this process to distinguish different new
physics models in future $THERA$ and $ILC$ experiments [28]. In this
paper, we have calculated the effective cross sections of the mirror
quark $Q_{H}$ production association with mirror neutrino $\nu_{H}$
via the subprocess $eq \rightarrow \nu_{H}Q_{H}$ at the $THERA$
experiment with $\sqrt{S}=3.7TeV$ and the $ILC$ experiment with
$\sqrt{S}=3TeV$. Our numerical results show that the values of the
cross sections are strongly dependent on the Yukawa coupling
parameter $k$ and the scale parameter $f$. This is because the
masses of the mirror quarks are written in a unified manner as
$M_{Q_{H}}=\sqrt{2}kf$. However, in wide range of the parameter
space, the mirror fermions can be abundantly produced.

Based on calculating the branching ratios of all possible two-body
decay modes of the mirror quarks $U_{H}$ and $D_{H}$, we further
calculate the production rates of certain signal events relevant for
the main decay modes. For the mirror quark decays to an ordinary
fermion and a T-odd gauge boson $B_{H}$, the subprocess $eq
\rightarrow \nu_{H}Q_{H}$ can induce the $jet$+$E \hspace{-0.25cm}
/$ signal event, the values of its statistical significance $SS$ in
the $THERA$ and $ILC$ experiments are given in Fig.5 and Fig.9,
respectively. As shown, in reasonable ranges of the free parameters
, their values can be significantly large. The possible signals of
the mirror quarks might be detected via this kind of decay channels
in future $THERA$ and $ILC$ experiments. We also calculate the
production rates of the signal events $E \hspace{-0.25cm} /$+$jets$
and $E \hspace{-0.25cm} /$ +$\ell+jet$, which come from the decay
channels $Q_{H} \rightarrow qZ_{H}$ and $Q_{H} \rightarrow qW_{H}$,
respectively. One can see from the relevant figures, for certain
ranges of the free parameters, there will generally be several tens
and up to thousands signal events to be generated per year. However,
the $SM$ backgrounds must be further studied.

\noindent{\bf Acknowledgments}

This work was supported in part by the National Natural Science
Foundation of China under Grants No.10675057 and Foundation of
Liaoning  Educational Committee(2007T086).

\end{document}